# Grid Databases for Shared Image Analysis in the MammoGrid Project

S. R. Amendolia[1], F. Estrella[1,2], T. Hauer[2], D. Manset[1,2], R. McClatchey[2], M. Odeh[2], T. Reading[3], D. Rogulin[1,2], D. Schottlander[3], T. Solomonides[1]

[1]ETT Division, CERN, 1211 Geneva 23, Switzerland
Email: Salvator.Amendolia@cern.ch

[2]CCCS Research Centre, Univ. of West of England, Frenchay, Bristol BS16 1QY, UK
Email:Richard.McClatchey@uwe.ac.uk

[3]Mirada Solutions Limited, Mill Street, Oxford, OX2 0JX, UK
Email:David.Schottlander@mirada-solutions.com

**Abstract**

*The MammoGrid project aims to prove that Grid infrastructures can be used for collaborative clinical analysis of database-resident but geographically distributed medical images. This requires: a) the provision of a clinician-facing front-end workstation and b) the ability to service real-world clinician queries across a distributed and federated database. The MammoGrid project will prove the viability of the Grid by harnessing its power to enable radiologists from geographically dispersed hospitals to share standardized mammograms, to compare diagnoses (with and without computer aided detection of tumours) and to perform sophisticated epidemiological studies across national boundaries. This paper outlines the approach taken in MammoGrid to seamlessly connect radiologist workstations across a Grid using an "information infrastructure" and a DICOM-compliant object model residing in multiple distributed data stores in Italy and the UK.*

## 1. Image Management and Analysis

Medical diagnosis and intervention increasingly relies upon images, of which there is a growing range available to the clinician: x-ray (increasingly digital, though still overwhelmingly film-based), ultrasound, MRI, CT, PET scans etc.

This trend will increase as high bandwidth (PACS) systems are installed in large numbers of hospitals (currently, primarily in large teaching hospitals).Patient management (diagnosis, treatment, continuing care, post-treatment assessment) is rarely straightforward; but there are a number of factors that make patient management based on medical images particularly difficult. Often very large quantities of data, with complex structure, are involved (such as 3-D images, time sequences, multiple imaging protocols). In most cases, no single imaging modality suffices, since there are many parameters that affect the appearance of an image and because clinically and epidemiologically significant signs are subtle including patient age, diet, lifestyle and clinical history, image acquisition parameters, and anatomical/ physiological variations.

To enable analysis of medical images related personal and clinical information (e.g. age, gender, disease status) have to be identified. The number of parameters that affect the appearance of an image is so large that the database of images developed at any single site - no matter how large - is unlikely to contain a set of exemplars in response to any given query that is statistically significant. Overcoming this problem implies constructing a huge, multi-centre - federated - database, while overcoming statistical biases such as lifestyle and diet leads to a database that may transcend national boundaries. For any medical condition, there would be huge gains if one had a pan-national database - so long as that (federated) database appears to the user as if it were installed in a single site. Such a geographically distributed (pan-European) database can be implemented using so-called Grid technology [1], and the construction of a prototype would enable a study of the suitability of Grid technologies for distributed mammogram analysis.

This paper outlines the advances made in the MammoGrid [2] project towards providing a collaborative Grid database analysis platform in which statistically significant sets of mammograms can be shared between clinicians across Europe. In the next section some important MammoGrid project objectives are identified and the role of the Information Infrastructure is highlighted. Then the essential underlying technologies on which this infrastructure is based are described. The MammoGrid Object Model is outlined in Section 3 including the DICOM Information Model and the so-called Assessment Object Model. The MammoGrid workstation application interface and how it maps onto a Grid infrastructure is then described prior to a discussion being undertaken on Grid query resolution before conclusions are drawn in the final section.

## 2. The MammoGrid Solution

### 2.1 Objectives

Amongst the objectives of the MammoGrid project are the need:

- To evaluate current Grid technologies and determine the requirements for Grid-compliance in a pan-European mammography database.
- To implement a prototype MammoGrid database, using novel Grid-compliant and federated-database technologies that will provide improved access to distributed data.
- To deploy versions of a standardization system (SMF - the Standard MammoGram Form [3, 4]) that enables comparison of mammograms in terms of tissue properties independently of scanner settings, and to explore its place in the context of medical image formats (e.g DICOM [5]) and
- To use the annotated information and the images in the database to benchmark the performance of the prototype system.

The MammoGrid project is being driven by the requirements of its user community (represented by Udine and Cambridge University hospitals along with medical imaging expertise in Oxford).

### 2.2 The Information Infrastructure

One of the main deliverables of the MammoGrid project is to provide an interface between a radiologists image analysis workstation and an 'MammoGrid Information Infrastructure' (MII) based on the philosophy of a Grid. This will enable radiologists to query images across a widely distributed federated database of mammographic images and to perform epidemiological and Computer Aided Detection CADe [6] analyses on the sets of returned images.

In delivering the MII the MammoGrid project is customising and, where necessary, enhancing and complementing Grid software for the creation of a pan-European medical analysis platform. It is not the intention of this project to produce Grid-specific middleware but rather to develop solutions for the medical practitioner using, where appropriate, solutions from the DataGrid [7] (and other Grid) projects. In other words although principally addressing the application needs of the clinician/radiologist, the MammoGrid project will incorporate new developments in its Grid infrastructure as and when those technologies become readily available and stable.

Current distributed computing technologies such as CORBA and Enterprise Java enable resource sharing within a single virtual organization (VO). The Open Group's Distributed Computing Environment (DCE) supports secure resource sharing across multiple sites, but most VOs find DCE too unwieldy and inflexible. In other words, current distributed technology either does not address the wide range of resources types or does not provide sufficient flexibility and the control needed for Grid-based VOs. This is the main reason why research communities are actively focusing on Grid technologies in order to enable heterogeneous resource sharing across multiple VOs.

This implies that the MII architecture must rely heavily on emerging Grid standards such as the Open Grid Services Architecture (OGSA [8]) and the Open Grid Services Infrastructure (OGSI [9]) and to have clearly delineated Application Program Interfaces (APIs) to software and services external to the MII. The approach that is being followed in MammoGrid is therefore two-fold: to provide an MII based on a service-oriented architecture with an OGSA-compliant gateway to multiple Grid implementations, and a meta-data and query handler coupled to a DICOM-server front-end so as to ensure both that data and images remain appropriately associated and that meta-data based searches are effectively handled (see [10]). The MII has been fully specified in MammoGrid and is being delivered in a set of staged prototypes in which a set of medical imaging services are implemented on an OGSA-compliant Grid infrastructure.

### 2.3 MammoGrid Technologies

**2.3.1 Introduction.** The MII, which federates multiple mammogram databases, will enable clinicians to develop new common, collaborative and cooperative approaches to the analysis of mammographic data. The following sections introduce the technologies used in integrating the MII with the radiologist workstation (provided by Mirada Solutions, Oxford, UK). Further detail are accessible through the MammoGrid's web pages [11].

**2.3.2 DICOM** (Digital Imaging & Communications in

Medicine) [5] is a widely used standard that addresses the exchange of digital information between medical imaging equipment and other systems . It covers storage with respect to interchange media devices such as writeable CDs and certain DVDs. The coverage of file formats within the standard relates to the syntax and semantics of commands and associated information which can be exchanged between devices using the protocols described in the standard, and to facilitate access to the images and related information stored on the interchange media.

The MammoGrid project aims to conform to the DICOM standard in two ways.  First, the digitized images should be imported and stored in the DICOM storage format (as DICOM files), so that the full set of image- and patient-related metadata is readily available with the images, and that information exchange with other medical devices understanding the DICOM storage format is seamless. To further ensure the compatibility with DICOM conformant clients, it is required that the exchange of DICOM datasets should be done via a communication protocol - also defined by the standard.  In this setup a client, or Service Class User (SCU) initiates a network connection with a server or Service Class Provider (SCP) and they exchange DICOM datasets over the established association protocol.

To facilitate DICOM compliance it is required that the server side (Grid-Box) exposes a DICOM SCP which is capable of establishing an association with an SCU started by the client side (Mirada workstation).  Mammograms should be transferred to the server for addition to the database via this association and similarly requested image files are expected to be ready for download via the SCU/SCP pair.

**2.3.3 AliEn and the 'Grid-Boxes'.** AliEn (Alice Environment) [12] is a Grid framework developed to satisfy the needs of the ALICE experiment at CERN for large scale distributed computing. It is built on top of the latest Internet standards for information exchange and authentication (SOAP, SASL, PKI) and common Open Source components (such as Globus/GSI, OpenSSL, OpenLDAP, SOAPLite, MySQL, CPAN). AliEn provides a virtual file catalogue that allows transparent access to distributed datasets and at the same time, AliEn provides an insulation layer between different Grid implementations and provides a stable user and application interface to the community of Alice users during the expected lifetime of the experiment. As progress is being made in the definition of Grid standards and interoperability, AliEn will be progressively interfaced to the DataGrid [7] as well as to other Grid structures.

The CERN AliEn software has been installed and configured on a set of novel 'Grid-Boxes', or secure hardware units, which will act as each hospital's single point of entry onto the MammoGrid and will provide the security and control of access needed for sensitive medical data. These units are being configured and tested at CERN and Oxford, for later testing and integration with other Grid-Boxes in the Udine and Cambridge hospitals. Each hospital has direct, secure access to a dedicated Grid-Box via its local area network. Each Grid-Box can be seen as a "gate" to the Grid and is in charge of storing new Patient images / studies, updating the file catalogue and propagating the changes. The synchronous operation of the net of Grid-Boxes ensures that the view of database at the workstations is up to date at every site. The data sent through this network is anonymized and encrypted - an essential requirement for security and confidentiality.

While the Grid provides part of the essential features for the MammoGrid project, it is an essential security and confidentiality requirement that users should not interact directly with grid functionalities.  Rather, it is required that access to the mammogram database is exposed through a workstation client with a custom-made user interface. The first MammoGrid prototype uses a default client - a workstation developed by one of the partners (Mirada Solutions) - but the interface design aims to be general enough so that possibly other clients can be accommodated. This is achieved in three ways (1) the use of the industry-standard DICOM protocol for exchanging digital image and image-related data (2) the use of a standard-communication protocol model (SOAP and web services) for decentralized, distributed environment and (3) the use of the W3C XML/XSD for data exchange formats.

## 3. The MammoGrid Object Model

This section introduces the MammoGrid Object Model (MOM) and briefly describes its structure and enhancements to the DICOM Information Model (DIM).

### 3.1 The MOM and the DIM

The MOM is an object model which stores and manipulates data from DICOM files and provides the basis for the interface between the radiologists workstation and the Grids information infrastructure. In addition, it provides the core functionality for storing, querying and manipulating digital image data. Mammograms are kept in the DICOM file in the form of binary pixel data with the associated image-related metadata (e.g., acquisition and positioning information) as well as patient-related metadata, (e.g., gender, age, analysis performed, diagnosis etc.).

The DICOM Information Model (DIM) defines the structure and organization of the information related to the communication of medical images. It is used to model the relationships between 'real-world objects' which are defined in the DICOM Standard. The DIM model represents the hierarchy of the real world objects (in the familiar DICOM Patient-Study-Series-Image structure) and maps it in the way images are collected and managed. At the patient level, the identification and demographic information about the patient is handled. The result of a request for a certain type of examination is kept at the study level. The Series level identifies the modality type, details about examination and equipment used. The Image level contains acquisition and positioning information as well as the image data itself.

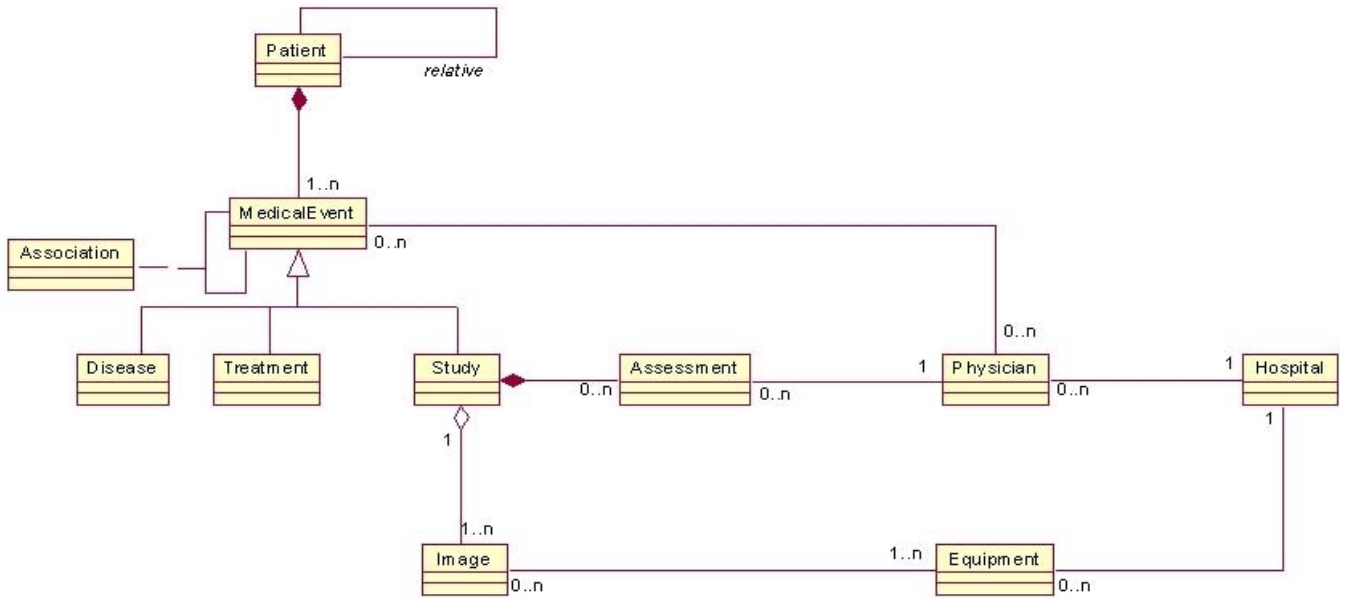

**Figure 1:** The MammoGrid Object Model

Although this model allows for storing and manipulating the image-related information it is insufficiently rich to provide clinicians with powerful, extensible and effective representation of mammogram data. In this project we are therefore augmenting the DIM with the MOM.

### 3.2 Structure of the MOM

The MammoGrid Object Model (MOM) extends the DIM with elements required for the support of clinician query resolution. It is a general model for medical imaging applications which provides the core functionality for storing, querying and manipulating digital image data. Figure 1 illustrates the high level representation of the MOM.

Conceptually the model is based around a set of Medical Event objects where, for example, a Medical Event may be a Patient's visit to a doctor or a Physician's interpretation of an Image. Medical Events can be inter-related and there can be dependences or complex associations between them (for example, to record things like "Drug Treatment as a consequence of Disease"). Doctors (physicians) make observations of Images and produce Assessments (often in the form of annotations) as the result of the examinations. These assessments are related to a study (as in the DICOM sense of a study) and indirectly to the medical images.

The MOM therefore is a model which not only represents DICOM information but also a framework which can be extended, using modeling "hooks", to cater for other digital medical image formats (e.g. Papyrus, Interfile etc.). This is the first time that such a DICOM-compliant model has been developed for Grids applications and this represents an important breakthrough in the use of Grids technologies for medical applications since DICOM has widespread acceptance in the field of Medical Informatics.

### 3.3 Annotation

In image analysis radiologists make observations on patients, assess the medical status of the patient and draw conclusions, make suggestions and perform medical procedures based on this knowledge. They also exchange medical information with each other, a process of essential importance since any patient today may be examined and treated by more than one person and because the accumulated knowledge has to be transferred for educational purposes. One important task is to find a suitable representation of the data which is retained in the database, which is sufficiently structured so that clinical queries can be run and at the same time captures as much as possible from the physicians' diagnostic description.

The following three-layer view is useful for the analysis of this domain:

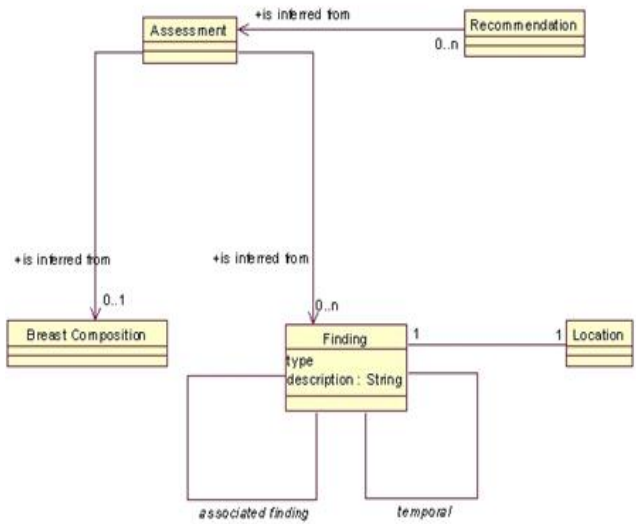

**Figure 2 :** The Assessment Object Model

1. Pathology as a real world notion. The identification and understanding of the underlying pathology, such as a particular lesion, is clearly useful for developing a successful reporting scheme.

2. Assessment of the findings by a physician. The radiologist has acquired information about the patient, obtained from many sources which is analysed based on previous experience, knowledge, etc. The radiologist ultimately aims to describe this assessment in the pathology report.

3. The Pathology Report finally is a description of the findings. There are multiple reporting methods that can be usefully incorporated into any model of a reporting scheme.

This three-layer approach is useful because a clear distinction can be maintained between real world objects (pathologies), what is being described and how it is being described. Figure 2 represents elements of the so-called Assessment Object Model (AOM) that is being used in MammoGrid to cater for medical findings, assessments and recommendations. The overall composition of the breast is defined along with an interpretation of the findings at the particular locations and a radiologist's assessment and recommendation. The AOM is an important element of the MOM. While there is only one real world object, there may be many assessments: different radiologists, different times, equipments, etc.

## 4. The MammoGrid Workstation Application Interface

### 4.1 Simplified Grid Infrastructure Description

Figure 3 shows a simplified view of the grid infrastructure proposed for MammoGrid. At each local site there is a local grid server; a Linux server connected via high speed Ethernet to the other Grid-boxes and acting as a gateway to the grid. The Grid-box is also connected to the local site Ethernet via a Gigabyte connection. All local workstations (Mirada WST (MAS)) that require grid connectivity do so via the Grid-box. In the simple model presented, the Grid is comprised of the sum of connected grid servers and potentially, a central server for administration tasks and centralised database functions.

In the model presented in Figure 3, the core data consists of DICOM files. Each hospital stores its own files on its local Grid-box maintaining ownership and responsibility for its data. The meta-data is extracted by grid processing when the file is stored in the relational database of the site's Grid server for efficient query processing.

The key point to note is that the method of interaction between workstation and grid must be abstracted from the grid deployment model through a common API. The next section describes the flow of data across the interface between the workstation and the Grid-box holding the MammoGrid Object Model.

### 4.2 Data Flows Between Grid Nodes and the Clinician Workstation

There are two types of data that flow between grid nodes and the workstation: Images and Data, represented by DICOM file exchange and Data only, represented by pure method invocation via a Soap API. In-line with the DICOM standard, it is considered best practice to transmit all required patient data as part-and-parcel of the file that contains the actual image under consideration to ensure integrity and completeness of the data. Principally for this reason it is proposed that network file exchange for image files between grid and workstation shall be DICOM3 conformant, using DICOM C-STORE and C-GET methods for image storage and retrieval respectively. DICOM C-FIND is however considered overly restrictive for the complex queries envisaged and hence a custom SOAP based query mechanism will be implemented.

Figure 4 presents a diagrammatic view of the data flow for two core processes, Acquire New Image and a Simple Query/Retrieve. These are discussed in the next sections

**4.2.1 Case 1 - Acquire New Image.** In this instance, the workstation acquires a DICOM file representing patient information and a radiological image and sends the file to the local grid server via the DICOM SCU/SCP pair. The Grid-box then saves the file locally (as an immutable object) and extracts the meta-information from the file. This is then sent to the database for storage along with any other non-image/patient related information such as log information, audit trail etc.

**4.2.2 Case 2 - Simple Query/Retrieve.** This is a much

more complex scenario although its presentation is still somewhat simplified for the sake of clarity. Here, the workstation starts the transaction by sending a query containing search criteria to the grid server. The grid server executes the search against the database (wherever it is located) and returns the set of Logical File Names (LFNs) and limited descriptive information that match the search criteria to the workstation.

The next transaction that occurs is when the user selects a particular case or set of cases to retrieve and display. The workstation then sends a request to the Grid-box with a list of LFNs relating to patient cases that it will require. This will trigger the Grid-box to start retrieving the cases to the local Grid-box cache in readiness for the workstation requesting the actual image files. The response from the Grid-box to this statement is the same list of LFNs sorted by estimated time required to access the file. In other words, cases where all the files are available on the local cache will be sorted on the top of the list.

The workstation will next issue a retrieve request to the Grid-box for each file it wishes to retrieve individually. The Grid-box will respond in one of three ways:
- If the file is currently being transferred from a remote Grid-box to the local cache, the request will block until the file is available locally and then the Physical File Name (the PFN to a dicom server, port and application entity address, such as DICOM:// ipaddress:port:aetitle:sopInstanceUid) is requested .
- If the file is available in the local cache, the PFN to the local DICOM server and file will be immediately returned
- If the file is available remotely, the PFN to the remote DICOM server and file will be immediately returned. In this instance, the file will still be retrieved to the local Grid-box cache as the likelihood is that the file will be required again in the near future.

In all cases, the workstation will receive a PFN which it will use to generate an instruction to a DICOM SCU to retrieve the actual DICOM file over the SCU/SCP C_GET network protocol.

**4.2.3 Case 3 - Updated Patient Details and Annotations.**
*4.2.3a Modify Patient Meta-Data.* A set of patient files (with or without image data) are retrieved for modification. Some patient information is changed and an update request is sent via the API to the grid server. This data is saved directly by the grid server on the database. The DICOM files for the patient are NOT modified or replaced at this point.

*4.2.3b Annotation, Classification or Storage of CADe Results.* A set of patient files are retrieved with image data as outlined in Case 2. All the data related to forming an opinion about a patient case, i.e. classification data, image annotations (contours, marks etc) and CADe markings arises from study of an image or set of images. This is then structured as a separate file or set of files associated with the patient, study, series or particular image. These files are conformant with the DICOM Structured Reporting specification and in particular the Mammography CADe SR IOD Templates defined in part 3.16 of the DICOM standard. The SR files are then sent to the grid server for storage and are then handled as per ordinary DICOM files described in Case 1.

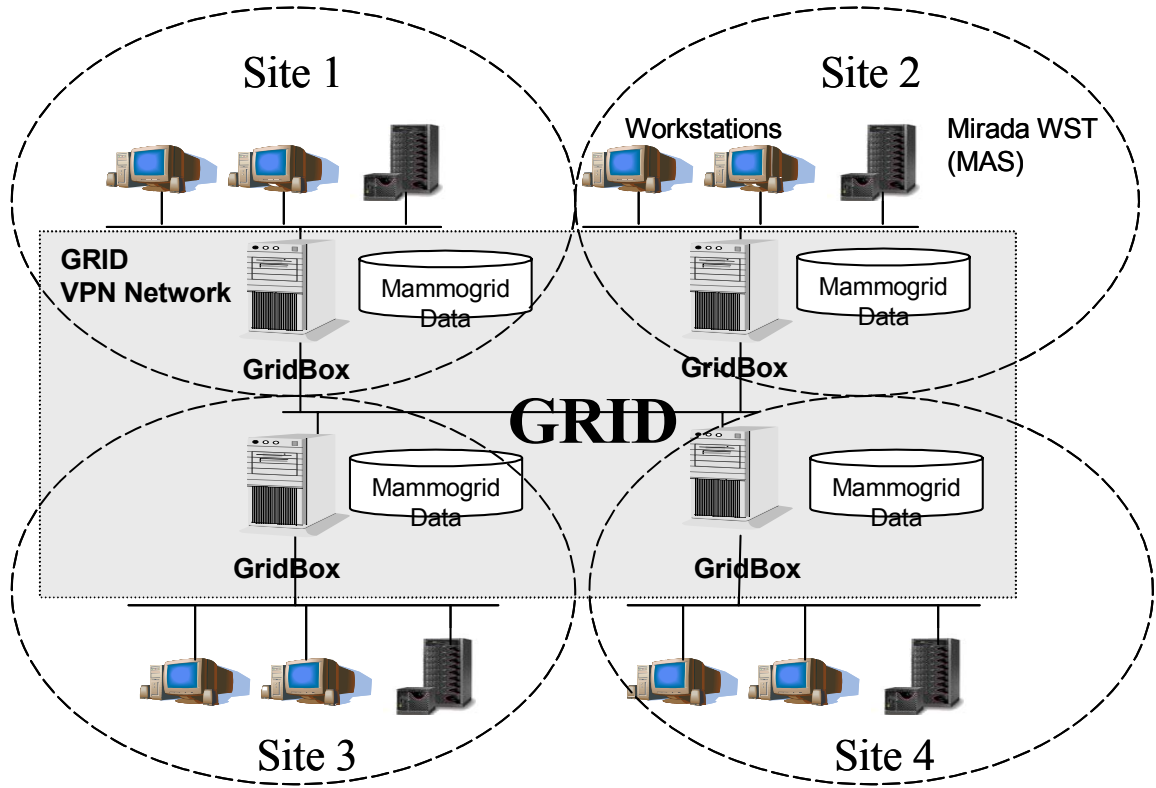

**Figure 3:** Simplified Grid Infrastructure for MammoGrid.

*4.2.3c Query/Retrieve with Modifications Applied to the Meta-Data.* The Query/Retrieve mechanism here from the point of view of the workstation operates identically to that described in Case 2. The difference occurs on the Grid-box. Once all the data is available locally, the original DICOM file must be merged with the modifications from the database, supplemented with the DICOM SR files and the resultant file-set stored for the interim on the local Grid-box cache. Only once this has been done, can the newly defined PFNs be returned to the workstation for referencing. Clearly, the only difference apparent to the workstation is that it must be able to support receipt of DICOM SR files.

**4.2.4 Complex Query/Retrieve.** The final case presented is a combination of the other cases. Considering applications such as 'Find-One-Like-It' there is a requirement to be able to perform queries based on image annotations, contours, classifications etc. In all probability, searches like this should use existing annotations already stored in SR files, however in certain searches it is possible that new contours may be drawn for search. In either case, the query mechanism needs to be able to take a structure representing an SR file or a LFN referencing an SR file in the search criteria.

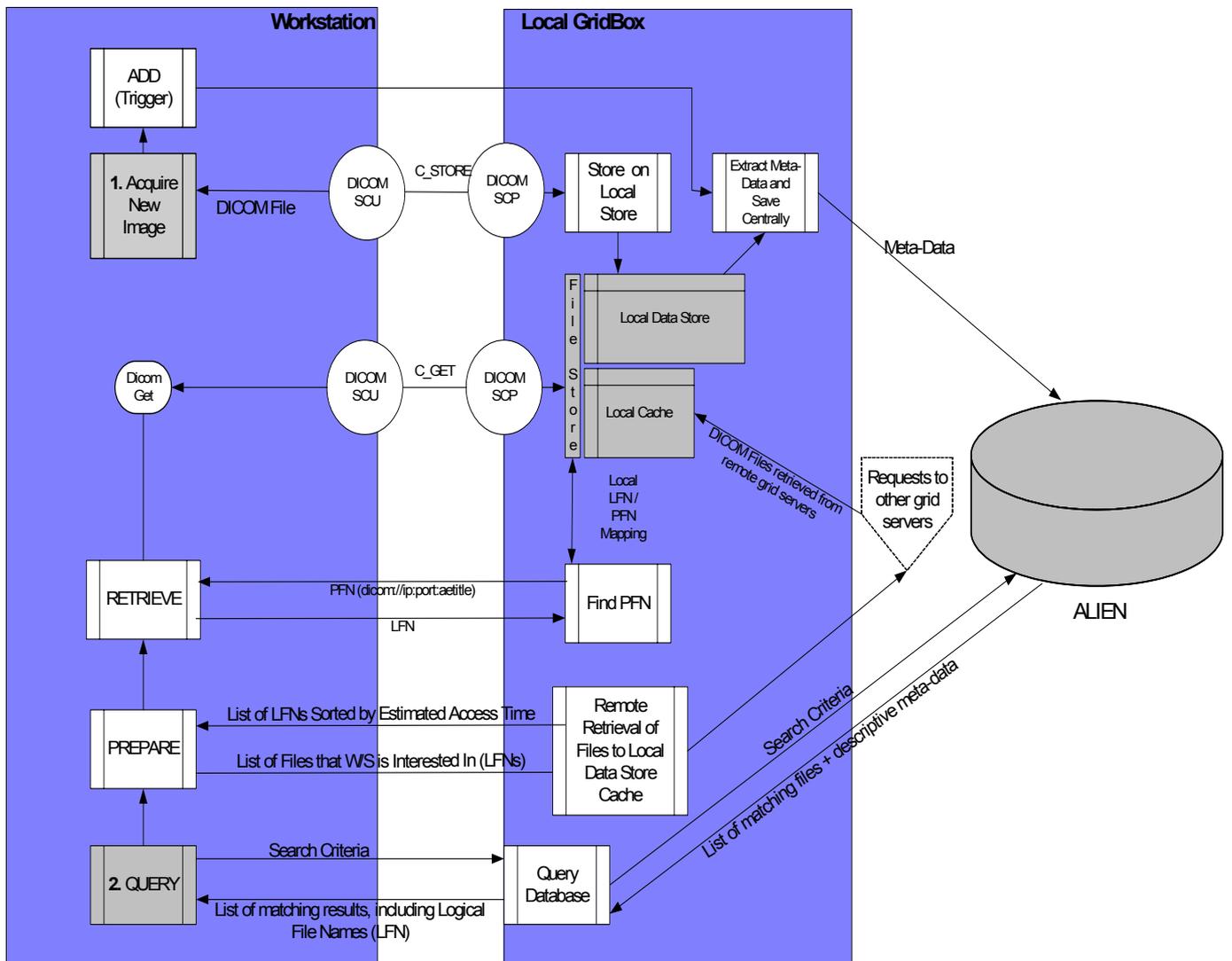

**Figure 4:** Data-flow between the radiologist workstation and the Grid-Box

# 5. Generalisation of the Architecture to Support Local & Distributed Data-Stores

By constraining all exchanges of data to be DICOM file exchange and implementing a limited API for managing query/retrieve functions, the implementation of the grid server is made independent of the workstation. In light of this, so long as the API is described using open standards it is possible to seamlessly exchange the grid server with any other data store mechanism that implements the interface.

This generalisation then takes the form of building on standards that support loose coupling and coarse-grained connection of distributed components. That is, to use XML [13] for description of meta-data and SOAP [14] for remote connectivity. XML offers a non-proprietary (non-binary) format that eliminates any networking, operating system or platform binding that a protocol has. It may seem contradictory to have stated previously that the image data will be exchanged using DICOM protocols whilst the meta-data for non-image related queries will be XML based - however taken in the context of the medical industry, there has been convergence on DICOM as the format for medical image exchange whilst the issue of complex distributed data management and tele-radiology is not addressed within the standard. SOAP provides a standard packaging structure for transporting XML documents over a variety of standard Internet technologies, thus it provides a medium for connectivity bypassing security issues related to firewalls and offering substantial flexibility in respect of deployment strategies.

This, in a nutshell, describes the essence of a Web Services architecture and in fact, there is increasingly convergence in the 'grid world' on developing OGSA [9], a Web Services interface to grid services. Thus, we intend that the communication between the workstation and the grid server, the workstation and the local data-store and the workstation and the remote data-store will all be conducted through the interaction of the client workstation with grid-services that expose the same interface, irrespective of platform or back-end implementation.

Medical conditions such as breast cancer, and mammograms as images, are extremely complex with many dimensions of variability across the population. Similarly, the way diagnostic systems are used and maintained by clinicians varies between imaging centres and breast screening programmes, and in consequence so does the appearance of the mammograms generated. An effective solution for the management of disparate mammogram data sources is a federation of autonomous multi-centre sites which transcends national boundaries. This is achieved by the creation of so-called *virtual organisations* (VOs) within the Grid and can be handled in two different ways, either the Grid is composed of a single VO which federates the different sites or it is composed of multiple VOs according to inter-site security agreements. Having multiple VOs or not has a direct impact on both data security and privacy and also on query complexity. These are discussed below.

## 5.1 Federation in a Single Virtual Organisation

In a single VO the resources in the MammoGrid federation, i.e. hospitals, research institutes and universities are governed by the same sharing rules with respect to authentication, authorization, resource and data access. These rules create a highly controlled environment which dictates what data are shared, who is allowed to share, and the conditions under which sharing occurs among members of the federation. Federation in this application implies cooperation of independent medical sites. Individually, these sites are autonomous in that they have separate and independent control of their local data. Collectively, these sites participate in a federation, and the federation is governed by the virtual organization.

In the current MammoGrid prototype the AliEn middleware provides services (e.g. authentication, data access, resource broker, file transfer) that facilitate the management of resources in the VO. In essence, the medical community dictates the interaction protocol, and AliEn implements and enforces these rules on the participating entities of the organization through services.

## 5.2 Federation in Multiple Virtual Organisations

The medical sites in a single VO operate within the rules specified by a governing organization. In reality, there are many (co)-existing organizations, with different rules and protocols. Typically, hospitals have different regulations and governments have different legislations. A federation of multiple VOs extends the single VO setup by inter-connecting potentially disparate VOs (see Figure 5). However, this VO mechanism is not sufficient to guarantee security and privacy. Some technical additions are needed to meet security requierements. Thus, two security aspects must be considered, first data encryption across Grid communications and second user authentication and authorization across VOs.

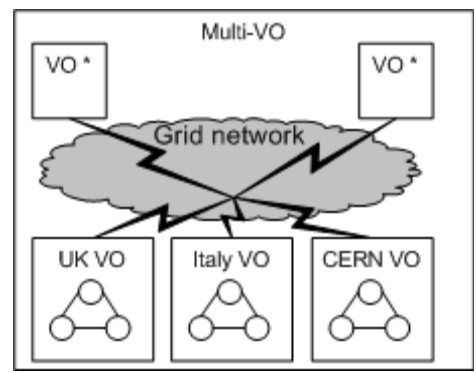

**Figure 5:** Federation in AliEn Multiple VOs

In order to preserve privacy, patient personal data is - as a first step - partially encrypted in MammoGrid to facilitate anonymization. As a second step, when data is transferred from one service to another through the network, communications are established through a secured protocol HTTPS with encryption at a lower level. Each Grid-box is in fact made part of a VPN (Virtual Private Network) in which allowed participants are identified by an unique host certificate.

AliEn provides a single sign-on mechanism based on the PKI Security model. This mechanism, in addition to the VO management service, enables the user to navigate through the allowed VOs. An unique certificate - delivered by the user's organization - is assigned to each user of the system. When the user is authenticated to MammoGrid, web services are used to interact within the Grid on behalf of the user using the user's credentials.

## 6. Query Resolution Across the Grid

In the MammoGrid proof-of-concept demonstrator, real clinician queries will be handled and resolved against data resident across a Grids infrastructure. User Requirements have been gathered that will enable queries to be executed and data retrieved for the analysis of mammograms. In particular the MammoGrid project will test the access to sets of mammogram images for the purposes of breast density assessment and for the testing of CADe studies of mammograms.

Queries can be categorized into simple and complex queries. Simple queries use predicates that refer to simple attributes of meta-data saved alongside the mammographic images. One example of a simple query might be to 'find all mammograms for women aged between 50 and 55' or 'find all mammograms for all women over 50 undergoing HRT treatment'. Provided that age and HRT related data is stored for (at least a subset of) patients in the patient meta-data then it is relatively simple to select the candidate images from the complete set of images either in one location of across multiple locations. It is also possible to collect data concerning availability of requested items so as to inform the design of future protocols, thus engineering a built-in enhancement process.

There are, however, queries which refer to data that has not been stored as simple attributes in the meta-data but rather require derived data to be interrogated or an algorithm to be executed. Examples of these might be queries that refer to the semi-structured data stored with the images through annotation or clinician diagnosis or that is returned by, for example, the execution of the CADe image algorithms.

During the final phase of implementation and testing, lasting until the completion of the project, the meta-data structures required to resolve the clinicians' queries will be delivered using the meta-modelling concepts of the CRISTAL project [15], [16]. This will involve customizing a set of structures that will describe mammograms, their related medical annotations and the queries that can be issued against these data. The meta-data structures will be stored in a database at each node in the MammoGrid (e.g. at each hospital or medical centre) and will provide information on the content and usage of (sets of) mammograms.

The query handling tool will locally capture the elements of a clinician's query and will issue a query, using appropriate Grids software, against the meta-data structures held in the distributed hospitals. At each location the queries will be resolved against the meta-data and the constituent sub-queries will be remotely executed against the mammogram databases. The selected set of matching mammograms will then be either analyzed remotely or will be replicated back to the centre at which the clinician issued the query for subsequent local analysis, depending on the philosophy adopted in the underlying Grids software. All data objects will reside in standard commercial databases, which will also hold descriptions of the data items.

## 7. Related Work and Conclusions

Other work in this area includes the NDMA [17] project in the US and the eDiamond [18] project in the UK. Our approach shares many similarities, but in the case of the NDMA project (one of whose principal aims is to encourage the adoption of digital mammography in the USA) its database is implemented in IBM's DB2 on a single server - that is, it avoids the technical issues of constructing a distributed database that exploits the emerging potential of the Grid. The MammoGrid project federates multiple (potentially heterogeneous) databases as its data store(s). MammoGrid is complementary to eDiamond and addresses different objectives : MammoGrid concentrates on the use of open source Grid solutions to perform epidemiological and CADe studies and incorporates pan-european data whereas eDiamond uses IBM-supplied Grid solution to enable 'find-one-like-it' image and teaching studies on UK data samples.

The current status of MammoGrid is that a single 'virtual organisation' AliEn solution has been demonstrated using the MII and images have been accessed and transferred between hospitals in the UK and Italy. The next stage is to provide rich meta-data structures and a distributed database to enable epidemiological queries to be serviced and the implementation of a service-oriented (OGSA-compliant) architecture for the MII.

The proliferation of information technology in medical sciences will undoubtedly continue, addressing clinical demands and providing increasing functionality. The MammoGrid project aims to advance deep inside this territory and explore the requirements of evidence-based, computation-aided radiology, as specified by medical scientists and practicing clinicians. This paper has emphasized two aspects which are likely to prove essential to the success of such a project: the importance of extensive requirements analysis and a design which caters for the complexity of the data. Currently the MammoGrid project is undertaking the implementation and testing of a first prototype in which a reduced set of mammograms are being tested between sites in the UK, Switzerland and Italy. Clinicians are being closely involved with these tests and it is intended that a subset of the clinician queries listed in section 3 will be executed to solicit user feedback. Within the next year a rigorous evaluation of the prototype will then indicate the usefulness of the Grid as a platform for distributed mammogram analysis and in particular for resolving clinicans' queries.

In its first year, the MammoGrid project has faced interesting challenges originating from the interplay between medical and computer sciences and has witnessed the excitement of the user community whose expectations from the a new paradigm are understandably high. As the MammoGrid project moves into its final implementation and testing phase, further challenges are anticipated which will test these ideas to the fullIn conclusion, this paper has described the approach taken in MammoGrid to seamlessly connect radiologist workstations across a Grid using an "information infrastructure" and a DICOM-compliant object model residing in multiple, distributed data stores in Italy and the UK.

**Acknowledgements**

The authors take this opportunity to acknowledge the support of their institutes and numerous colleagues responsible for the MammoGrid user requirements. The contributions of Professor Massimo Bazzocchi, Dr Ruth Warren and Dr Chiara del Frate are particularly acknowledged.

**References**

[1] I. Foster, C. Kesselman & S. Tueke., "The Anatomy of the Grid - Enabling Scalable Virtual Organisations", *Int. Journal of Supercomputer Applications*, 15(3), 2001.

[2] The Information Societies Technology project: "MammoGrid - A European federated mammogram database implemented on a GRID infrastructure". EU Contract IST-2001-37614

[3] SMF : Mirada Solutions' Standard Mammogram Form. See http://www.mirada-solutions.com/smf.htm

[4] R. Highnam & M. Brady., "Mammographic Image Analysis". *Kluwer publishers*, 1999.

[5] The DICOM Standard. Digital Imaging and Communications in Medicine. See http://medical.nema.org/

[6] C.J. Viborny, M.L. Giger, R.M. Nishikawa, "Computer aided detection and diagnosis of breast cancer", *Radiol. Clin. N. Am*. 38(4), 725-740, 2000.

[7] The Information Societies Technology project: "DataGRID - Research and Tecnological Development for an International Data Grid". EU Contract IST-2000-25182 See http://eu-datagrid.web.cern.ch/eu-datagrid/

[8] I. Foster, C. Kesselman J. Nick & S. Tueke., "The Physiology of the Grid - An Open Services Grid Architecture for Distributed Systems Integration". Draft document at http://www.globus.org/research/papers/ogsa.pdf

[9] OGSI. The Open Grids Services Infrastructure. See http://www106.ibm.com/developerworks/grid/library/gr-gt3/

[10] MammoGrid User Requirements. Deliverable D2.1. and MammoGrid Technical Specification, Deliverable 3.1 See http://mammogrid.vitamib.com

[11] MammoGrid home page see http://mammogrid.vitamib.com/

[12] P. Saiz, L. Aphecetche, P. Buncic, R. Piskac, J. -E. Revsbech and V. Sego., "AliEn - ALICE environment on the GRID", *Nuclear Instruments and Methods A* 502 (2003) 437-440, and http://alien.cern.ch

[13] XML - The Extensible Markup Language. See http://www.w3.org/XML/

[14] SOAP - The Simple Object Access Protocol. See http://www.w3.org/TR/SOAP/

[15] F.Estrella, Z. Kovacs, J-M. Le Goff & R. McClatchey., "Meta-Data Objects as the Basis for System Evolution". *Lecture Notes in Computer Science* Vol 2118, pp 390-399 ISBN 3-540-42298-6 Springer-Verlag, 2001.

[16] F. Estrella, J-M Le Goff, Z. Kovacs, R McClatchey & S. Gaspard., "Promoting Reuse Through the Capture of System Description" *Lecture Notes in Computer Science* Vol 2426 p 101-111 ISBN 3-540-44088-7 Springer-Verlag, 2002.

[17] NDMA: The National Digital Mammography Archive. Contact Mitchell D. Schnall, M.D., Ph.D., University of Pennsylvania. See http://nscp01.physics.upenn.edu/ndma/projovw.htm

[18] M. Brady, M. Gavaghan, A. Simpson, M. Mulet-Parada & R. Highnam., "eDiamond: A Grid-enabled Federated Database of Annotated MammoGrams" paper in "Grid Computing: Making The Global Infrastructure a Reality", Eds. F. Berman, G. Fox & T. Hey. *Wiley publishers,* 2003.